\documentclass[twocolumn,pra,superscriptaddress,showpacs,aps,floatfix]{revtex4-1}
\usepackage{color}
\usepackage{amsmath}
\usepackage{amssymb}
\usepackage{txfonts}
\usepackage{graphicx}
\usepackage{epstopdf}
\usepackage[unicode]{hyperref}
\hypersetup{pdftitle={2D SOC BEC},
 pdfauthor={},
 pdfsubject={2D SOC BEC},
  colorlinks=true,
    linkcolor=red,
    citecolor=magenta,
    filecolor=black,
    urlcolor=blue}


\begin{document}

\title{The Two Higgs-Doublet Extension of the Standard Model and S3 Symmetry}

\author{Chilong Lin, ~Chien-Er Lee~and~Yeou-Wei Yang }
\email{lingo@mail.nmns.edu.tw}

\affiliation{Department of Physics, National Chien Kung University, Tainan, Taiwan}

\date{Published in Chin. J. of Phys., Vol. 26, No. 4, 180. Received September 14, 1988. Recompiled \today }

\begin{abstract}
We impose a permutation symmetry S3 upon the two Higgs doublet extension of the SU(2) x U( 1) standard model,
and then break S3 symmetry spontaneously.
A mass relation is obtained from which the mass of the top quark is predicted to be 39 Gev at 1 Gev renormalization scale,
if the relative strength between the coupling of the first Higgs field with the up-type quarks is assumed to be the same as
that with the down-type quarks.
We find that in our model the flavor changing neutral currents vanish automatically.
\end{abstract}
\maketitle


\section{Introduction}

It is generally believed that there are at least three generations of quarks. 
Among them only the top quark is still illusive experimentally.
A fundamental problem in particle physics is the generation of both the mass spectrum of the quarks and leptons and the related
weak-mixing matrices which represent the coupling strength of the weak charged currents.
In our earlier attack~\cite{Lee1986} on these problems,
we extended the permutation S3-group invariance of the standard SU(2) x U(1) Lagrangian into its Higgs sector and found it successful.
Unfortunately the S3 breaking term was however chosen quite artificially.
In this article, we let S3 symmetry break spontaneously; therefore S3 breaking ensues in a natural way. \\

In order to achieve this objective, we here consider the two Higgs-doublet extension of the standard model.
Many physicists ~\cite{TDLee1973} have suggested and investigated the two Higgs doublet models,
but they met the problem of the non-vanishing flavor-changing neutral currents~\cite{Branco1985} (F.C.N.C.) when they apply both Higgs doublets to the same type of quark fields.
Their method to avoid the F.C.N.C. problem was to couple the different Higgs doublets
separately to different types of quark. \\

In our model of the two Higgs-doublet extension of the standard model,
we impose the permutation S3 symmetry in the Lagrangian and then break it spontaneously. We find that
because the flavor changing neutral current is automatically forbidden, there is no F.C.N.C. problem in our model. \\

There are three irreducible representations of S3~\cite{Boardman1983}, denoted by $\Gamma^1$, $\Gamma^2$ and $\Gamma^3$ respectively.
In this article, we consider two different classifications for the two Higgs doubles $\Phi_1$ and $\Phi_2$:
(1) $\Phi_1$ belongs to the totally symmetric representation $\Gamma^1$ while $\Phi_2$ to the totally antisymmetric $\Gamma^2$;
(2)  $\Phi_1$ and $\Phi_2$ belong to the two-dimensional representation $\Gamma^3$.
As for the quarks, we classify them according to the three-dimensional reducible representation of S3 denoted by $\Gamma^6$ ,
which just corresponds to the obvious permutations on the quark generations.
The nonvanishing of the vacuum expectation values $< \Phi_1 >$ in case 1 and $< \Phi_1 >$ or $< \Phi_2> $ in case 2 then produces the spontaneous S3 breaking.
We find that the first case gives a reasonably good prediction of the mass of the top quark whereas the second case does not.
In the above classifications, no generation mixing shows up.
This lack suggests that more Higgs doublets~\cite{Frampton1985} or different classifications for right-handed quarks are needed. \\

In section 2, we impose S3 symmetry and obtain the relations among the coupling constants for case 1.
In section 3, the corresponding mass matrices are obtained, and the mass of the top quark is predicted.
In section 4, case 2 is considered.
 In the concluding section, we discuss our work.

\section{TWO HICGS-DOUBLET EXTENSION AND S3 SYMMETRY }
In this section, we consider the two Higgs-doublet extension of the standard model under the S3 symmetry. \\

We add another SU(2) x U(1) scalar doublet, $\Phi_2$, to the standard SU(2) x U( 1) model and let $\Phi_2$ have the same form as $\Phi_1$ in the standard model. \\
\begin{eqnarray}
\Phi_1 = \left( \begin{array}{c} \Phi_1^+   \\ \Phi_1^0 \end{array}\right),
~~\Phi_2 = \left( \begin{array}{c} \Phi_2^+   \\ \Phi_2^0 \end{array}\right). \nonumber ~~~~~~~(2.1)
\end{eqnarray}

As is shown in Appendix A, the vacuum expectation values can be written as
\begin{eqnarray}
 < \Phi_1 >= \left( \begin{array}{c} 0   \\ \rho_1 \end{array}\right)/ \sqrt{2},
~~< \Phi_2 >= \left( \begin{array}{c} 0   \\ \rho_2 \end{array}\right)/ e^{i\theta}\sqrt{2}.  \nonumber ~~~(2.2)
\end{eqnarray}
in which both $\rho_1$ and $\rho_2$ are assumed to be positive and $\theta$ is real.
That $\theta$ does not vanish implies that the time-reversal invariance is broken spontaneously. \\

We write the weak interaction Lagrangian as
\begin{eqnarray}
L(q', \Phi) &=&\bar{q'_L} (\Phi_1 g +\Phi_2 g') q'_{dR} +\bar{q'_L} (\tilde{\Phi_1} h +\tilde{\Phi_2} h' )q'_{uR} +H.C. \nonumber \\
  &=& \bar{q'_L} G q'_{dR} +\bar{q'_L} H q'_{uR} +H.C.  \nonumber ~~~~~~~(2.3)
\end{eqnarray}
in which

and g, g' (h, h') are the coupling constants of Higgs fields to the down- (up-)type quark fields. The $\tilde{\Phi_k}$ are given by
\begin{eqnarray}
\tilde{\Phi_k}=(-i \Phi_k^+ \cdot \tau_2)^T  \nonumber ~~~~~~~(2.6)
\end{eqnarray}

We assume that both $\tilde{\Phi_k}$ are invariant under time reversal, i.e, $T \Phi_k T^{-1} = \Phi_k$, k = 1, 2.
Then g, g' (h and h') must all be real if $L_{int}$ is T-invariant. \\
\begin{eqnarray}
G &=& \left( \begin{array}{ccc} 
g_{11} \Phi_1 + g'_{11} \Phi_2 & g_{12} \Phi_1 + g'_{12} \Phi_2 & g_{13} \Phi_1 + g'_{13} \Phi_2 \\
g_{21} \Phi_1 + g'_{21} \Phi_2 & g_{22} \Phi_1 + g'_{22} \Phi_2 & g_{23} \Phi_1 + g'_{23} \Phi_2 \\
g_{31} \Phi_1 + g'_{31} \Phi_2 & g_{32} \Phi_1 + g'_{32} \Phi_2 & g_{33} \Phi_1 + g'_{33} \Phi_2 \end{array}\right) , \nonumber ~~~(2.4) \\
H &=& \left( \begin{array}{ccc} 
h_{11} \tilde{\Phi_1} + h'_{11} \tilde{\Phi_2} & h_{12} \tilde{\Phi_1} + h'_{12} \tilde{\Phi_2} & h_{13} \tilde{\Phi_1} + h'_{13} \tilde{\Phi_2} \\
h_{21} \tilde{\Phi_1} + h'_{21} \tilde{\Phi_2} & h_{22} \tilde{\Phi_1} + h'_{22} \tilde{\Phi_2} & h_{23} \tilde{\Phi_1} + h'_{23} \tilde{\Phi_2} \\
h_{31} \tilde{\Phi_1} + h'_{31} \tilde{\Phi_2} & h_{32} \tilde{\Phi_1} + h'_{32} \tilde{\Phi_2} & h_{33} \tilde{\Phi_1} + h'_{33} \tilde{\Phi_2} \end{array}\right).
 \nonumber ~~~(2.5)
\end{eqnarray}
As the standard model is invariant under the permutation of generations except for the Higgs sector,
we simply extend S3 symmetry to the Higgs sector;
that is, we assume that the Lagrangian is S3-invariant.
We then study the properties of the coupling constants under S3 symmetry.
In this section, we consider only the case that $\Phi_1$ is S3 symmetric (not changing sign under transposition $P_{ij}$)
 and $\Phi_2$ is S3 totally antisymmetric (changing sign under transposition $P_{ij}$). \\

 We study the properties of G and H under S3 symmetry.
 For convenience, we divide G and H into $G'$, $G"$ and $H'$, $H"$ separately, so that $G'$ and $H'$ correspond to $\Phi_1$
 and $G"$ and $H"$ to $\Phi_2$. \\

 (1) Under the transposition $P_{ij}$:
 \begin{eqnarray}
P_{23} G P_{23}^{-1} = \left( \begin{array}{ccc} G'_{11}-G"_{11} &  G'_{13}-G"_{13} & G'_{12}-G"_{12}   \\
G'_{31}-G"_{31} & G'_{33}-G"_{33} & G'_{32}-G"_{32} \\
G'_{21}-G"_{21} & G'_{23}-G"_{23} & G'_{22}-G"_{22} \end{array}\right) \nonumber ~~~(2.7) \\
P_{12} G P_{12}^{-1} = \left( \begin{array}{ccc} G'_{22}-G"_{22} &  G'_{21}-G"_{21} & G'_{23}-G"_{23}  \\
G'_{12}-G"_{12} & G'_{11}-G"_{11} & G'_{13}-G"_{13} \\
G'_{32}-G"_{32} & G'_{31}-G"_{31} & G'_{33}-G"_{33} \end{array}\right) \nonumber ~~~(2.9) \\
P_{13} G P_{13}^{-1} = \left( \begin{array}{ccc} G'_{33}-G"_{33} &  G'_{32}-G"_{32} & G'_{31}-G"_{31}  \\
G'_{23}-G"_{23} & G'_{22}-G"_{22} & G'_{21}-G"_{21} \\
G'_{13}-G"_{13} & G'_{12}-G"_{12} & G'_{11}-G"_{11} \end{array}\right) \nonumber ~~~(2.10)
\end{eqnarray}

(2) Under the cyclic permutation:
 \begin{eqnarray}
G_{123} {C_{231} \over  } G_{231} = \left( \begin{array}{ccc}
G'_{22}-G"_{22} & G'_{23}-G"_{23} & G'_{21}-G"_{21} \\
G'_{32}-G"_{32} & G'_{33}-G"_{33} & G'_{31}-G"_{31} \\
G'_{12}-G"_{12} & G'_{13}-G"_{13} & G'_{11}-G"_{11} \end{array}\right) \nonumber ~~~(2.11) \\
G_{123} {C_{312} \over } G_{312} = \left( \begin{array}{ccc}
G'_{33}-G"_{33} & G'_{31}-G"_{31} & G'_{32}-G"_{32} \\
G'_{13}-G"_{13} & G'_{11}-G"_{11} & G'_{12}-G"_{12} \\
G'_{23}-G"_{23} & G'_{21}-G"_{21} & G'_{22}-G"_{22} \end{array}\right) \nonumber ~~~(2.12) \\
\end{eqnarray}

Because we assume the Lagrangian density is S3-invariant, comparing the elements of the transformed matrices with the initial matrix, we find that
\begin{eqnarray}
G'_{11} &=& G'_{22} =G'_{33} \nonumber  \\
G'_{12} &=& G'_{21} =G'_{23} = G'_{32} = G'_{31} =G'_{13} \nonumber ~~~(2.13) \\
H'_{11} &=& H'_{22} =H'_{33} \nonumber \\
H'_{12} &=& H'_{21} =H'_{23} = H'_{32} = H'_{31} =H'_{13} \nonumber ~~~(2.14)
\end{eqnarray}
and
\begin{eqnarray}
G"_{ii} &=& 0,~~H"_{ii}=0 \nonumber ~~~~~~~~~(2.15) \\
G"_{21} &=& G"_{32} =G"_{13} = -G"_{23} = -G"_{31} =-G"_{12} \nonumber  \\
H"_{21} &=& H"_{32} =H"_{13} = -H"_{23} = -H"_{31} =-H"_{12} \nonumber ~~~(2.16)
\end{eqnarray}
which implies
\begin{eqnarray}
g_{ii} &=& a,~~h_{ii}=a' \nonumber ~~~~~~~(2.17) \\
g_{ij} &=& b,~~ h_{ij}=b'  \nonumber ~~~~~~~(2.18)  \\
g'_{ii} &=& 0, ~~h'_{ii}=0,~~i,j=1,2,3,~~if~ i\neq j \nonumber ~~~(2.19) \\
g'_{21} &=& g'_{32} =g'_{13} =-g'_{12} =-g'_{23} =-g'_{31}=c \nonumber ~~~(2.20) \\
h'_{21} &=& h'_{32} =h'_{13} =-h'_{12} =-h'_{23} =-h'_{31}=c' \nonumber ~~~(2.21) 
\end{eqnarray}

The matrices G and H then become
\begin{eqnarray}
G &=& \left( \begin{array}{ccc}
a \Phi_1 & b \Phi_1 -c\Phi_2 &  b \Phi_1 + c\Phi_2 \\
b \Phi_1 + c\Phi_2 & a \Phi_1 & b \Phi_1 -c\Phi_2 \\
b \Phi_1 -c\Phi_2 & b \Phi_1 + c\Phi_2 & a \Phi_1
\end{array}\right) , \nonumber ~~~(2.22) \\
H &=& \left( \begin{array}{ccc}
a' \tilde{\Phi_1}  &  b' \tilde{\Phi_1} - c' \tilde{\Phi_2} & b' \tilde{\Phi_1} + c' \tilde{\Phi_2} \\
b' \tilde{\Phi_1} + c' \tilde{\Phi_2} & a' \tilde{\Phi_1}  & b' \tilde{\Phi_1} - c' \tilde{\Phi_2} \\
b' \tilde{\Phi_1} - c' \tilde{\Phi_2} & b' \tilde{\Phi_1} + c' \tilde{\Phi_2} & a' \tilde{\Phi_1}  \end{array}\right) \nonumber ~~~(2.23) 
\end{eqnarray}

We obtain the mass matrices by substituting the vacuum expectation values of $\Phi_1$ and $\Phi_2$ into G and H.
We use the Hermitian basis~\cite{Chau1983} and let $A = a< \Phi_1^0 >,~ A' = a'< \Phi_1^0 >,~ B = b< \Phi_1^0 >,~ B' = b'< \Phi_1^0 >,~ C = c< \Phi_2^0 >$ and $C' = < \Phi_2^0 >$. Then under the Hermitian basis, we have
\begin{eqnarray}
A &=& a< \Phi_1^0 >=a< \Phi_1^0 >^* , \nonumber \\
B-C &=& b< \Phi_1^0 >-c< \Phi_2^0 > = b< \Phi_1^0 >^* + c< \Phi_2^0 >^*  \nonumber \\
A' &=& a'< \Phi_1^0 >=a'< \Phi_1^0 >^* , \nonumber ~~~(2.24)  \\
B'-C' &=& b'< \Phi_1^0 >-c'< \Phi_2^0 > = b'< \Phi_1^0 >^* + c'< \Phi_2^0 >^*
\end{eqnarray}

From (2.24), we easily see that A, A', B and B' are real and $C=i D$, $C'=i D'$ are purely imaginary.
We write the mass matrices Mu and Md as follows.
\begin{eqnarray}
M_d &=& \left( \begin{array}{ccc}
A  & B-iD &  B+i D \\
B+i D & A & B-i D \\
B-i D & B=i D & A
\end{array}\right) , \nonumber ~~~(2.25) \\
M_u &=& \left( \begin{array}{ccc}
A'  & B'-i D' &  B'+i D' \\
B'+i D' & A' & B'-i D' \\
B'-i D' & B'+i D' & A'   \end{array}\right) \nonumber ~~~(2.26) 
\end{eqnarray}

Because A, A', B, B', a, a', b, b', c and c' are all real, the vacuum expectation value of $\Phi_1$ must be real as required in Appendix A.
Conversely, because C and C' are purely imaginary, the vacuum expectation value of $\Phi_2$ must be imaginary.
That is, the phase angle (difference) $\theta$ can be only $\pi /2$ or $-\pi /2$. \\

We also note that, if both $\Phi_k$ are totally symmetric under S3 then the mass spectrum is degenerate.
If both $\Phi_k$ are totally antisymmetric under S3, there are negative eigenvalues in the mass spectrum.
As both events are contrary to nature, we abandon these two cases in this article. \\

\section{MASS SPECTRA AND THE WEAK MIXING MATRIX}

The mass matrices are diagonalized with unitary transformations Uu and Ud
\begin{eqnarray}
U_u M_u U_u^{-1} &=& diag.(m_u,~ m_c,~ m_t), \nonumber \\
U_d M_d U_d^{-1} &=& diag.(m_d,~ m_s,~ m_b) \nonumber ~~~(3.1) 
\end{eqnarray}

The eigenvalues of Mu and Md are solved and identified as
\begin{eqnarray}
m_d=A-B-\sqrt{3} D,~m_s=A-B+\sqrt{3} D,~m_b=A+2B, \nonumber (3.2)\\
m_u=A'-B'-\sqrt{3} D',~m_c=A'-B'+\sqrt{3} D',~m_t=A'+2B'. \nonumber (3.3)
\end{eqnarray}

It seems reasonable to assume that the relative strength between the coupling of $\Phi_1$ with the up-type quarks is the same as that with the down-type quarks. \\

Under this assumption we have
\begin{eqnarray}
a':b'=a:b ~~~~~~~\nonumber ~~~(3.4)
\end{eqnarray}

We then assign a proportionality factor L to this relation
\begin{eqnarray}
a'/a=b'/b=L ~~~~~~~\nonumber ~~~(3.5)
\end{eqnarray}
With this assignment, the parameters A, B, A' and B' in Md and Mu have the same relation
\begin{eqnarray}
A'/A=B'/B=L ~~~~~~~\nonumber ~~~(3.6)
\end{eqnarray}

From expressions (3.2) (3.3) and (3.6) we obtain the following relation
\begin{eqnarray}
m_t / m_b =(m_u +m_c) /(m_d+m_s) ~~~~~~~\nonumber ~~~(3.7)
\end{eqnarray}

We use the current quark masses which were renormalized at 1 Gev ~\cite{Gasser1982}: $M_d= 0.0089 Gev$, $m_s = 0.175 Gev$, $m_b = 5.3 Gev$, $m_u = 0.0051 Gev$ and $m_C = 1.35 Gev$. Then L =7.36868.

The mass of top quark is predicted to be
\begin{eqnarray}
m_t  =39.1 GeV ~~~~~~~\nonumber ~~~(3.8)
\end{eqnarray}
and the parameters are calculated to be: A = 1.82796 Gev, B = 1.73602 Gev, D = 0.04795 Gev, A'-B' = 0.67755 Gev, D' = 0.38824 Gev, A'= A $\times$ L = 13.46965 Gev, B'= B $\times$ L = 12.79218 Gev. \\

The transformation matrices Uu and Ud are found to have the same form and are independent of the parameters
\begin{eqnarray}
U_d=U_u = \left( \begin{array}{ccc}
1/\sqrt{3} & (-1-i \sqrt{3} /2\sqrt{3} & (-1+i \sqrt{3} /2\sqrt{3} \\
1/\sqrt{3} & (-1+i \sqrt{3} /2\sqrt{3} & (-1-i \sqrt{3} /2\sqrt{3} \\
1/\sqrt{3} & 1/\sqrt{3} & 1/\sqrt{3}
\end{array}\right) \nonumber (3.9)
\end{eqnarray}

Thus the weak mixing matrix defined by
\begin{eqnarray}
V=U_u \cdot U_d^{-1} ~~~~~~~\nonumber ~~~(3.10)
\end{eqnarray}
is a 3 $\times$ 3 identity matrix. That is, no CP-violation phase and mixing angles appear in this model.

In most of the two Higgs doublet models, there exist flavor-changing neutral current problems.
In these models, the different parts of the mass matrices correspond separately(respectively) to $\Phi_1$ and $\Phi_2$ could not be diagnoalized simultaneously.
When the quarks are transformed to the mass eigenstates, the coupling constant matrices of $\Phi_1$ and $\Phi_2$ are not diagonalized separately(respectively).
The off-diagonal parts prevent the flavor-changing neutral currents from vanishing but it is well known that the F.C.N.C. should be highly suppressed. \\

In our model, there exists no F.C.N.C. problem.
Those coupling constant matrices we have obtained can all be diagonalized by the same transformation matrix (3.9).
Therefore the flavor-changing neutral currents vanish automatically. \\

We also note that the current quark masses we use here have been renormalized at 1 Gev.
The results we state above should be corrected when they are renormalized at the scale $m_W$.
As for the $W$ and $Z^0$ masses in this model, we have only to replace the $< \Phi^0 >$ in the standard model by $\sqrt{<\Phi_1^0 >^2 +< \Phi_2^0 >^2}$.
With this replacement the $W$ and $Z^0$ masses are still the same as those in the standard model at the tree level. \\

\section{HIGGS DOUBLETS UNDER $\Gamma^3$ REPRESENTATIONS}

There are only three irreducible representations: $\Gamma^1$, $\Gamma^2$ and $\Gamma^3$ in S3 group.
$\Gamma^1$ and $\Gamma^2$ are one-dimensional totally symmetric and antisymmetric representations,
whereas $\Gamma^3$ is two-dimensional.
In sections (2) and (3), the two Higgs doublets are assumed to transform as $\Gamma^1$ and $\Gamma^2$ respectively,
whereas the quark fields with three generations transform as the three-dimensional representation $\Gamma^6$.
The $\Gamma^6$ representation is reducible and can be decomposed as $\Gamma^6$ = $\Gamma^1 \oplus \Gamma^3$.
 The representation $\Gamma^6 (R)$ with elements R = E, A, B, C, D and F are denoted as $\Gamma^6(E)$ = 1, $\Gamma^6(A)=P_{23}$,  $\Gamma^6(B)=P_{13}$, $\Gamma^6(C)=P_{12}$, $\Gamma^6(D)=C_{231}$, $\Gamma^6(F)=C_{312}$, in which $P_{ij}$ is the transposition and $C_{ijk}$ is the cyclic permutation. \\

For example
\begin{eqnarray}
\Gamma^6(A) q'_{dR}= \left( \begin{array}{ccc} 1 & 0 & 0 \\ 0 & 0 & 1 \\ 0 & 1 & 0 \end{array}\right)
\left( \begin{array}{ccc} d' \\ s' \\b' \end{array}\right)_R
=\left( \begin{array}{ccc} d' \\ b' \\ s' \end{array}\right)_R =P_{23} q'_{dR} \nonumber ~~~(4.1)
\end{eqnarray}
 as both $\bar{q'}$ and $q'$ transform as $\Gamma^6$.
 The $\bar{q'} q'$ parts in $L_{int}$ transform as the product representation $\Gamma^6 \otimes \Gamma^6$, which decomposes to
 \begin{eqnarray}
\Gamma^6 \otimes \Gamma^6= (\Gamma^1 \oplus \Gamma^3) \otimes (\Gamma^1 \oplus \Gamma^3)= 2\Gamma^1 \oplus \Gamma^2 \oplus 3\Gamma^3 \nonumber ~~~(4.2)
\end{eqnarray}
 in which we have used
  \begin{eqnarray}
\Gamma^1 \otimes \Gamma^3 &=& \Gamma^3 \otimes \Gamma^1 =\Gamma^3 \nonumber \\
\Gamma^3 \otimes \Gamma^3 &=& \Gamma^1 \oplus \Gamma^2 \oplus \Gamma^3 \nonumber ~~~~~~~~(4.3)
\end{eqnarray}

We find that $\sum_{i}^3 \bar{q'}_i q'_i$ and $\sum_{i \neq j}^3 \bar{q'}_i q'_j$ transform as $\Gamma^1$ and $\sum_{i,j,k}^3 \epsilon_{ijk} \bar{q'}_i q'_j$ transform as $\Gamma^2$.
Under the classifications in the previous sections, $L_{int}$ is written (as)
\begin{eqnarray}
L_{int} &=& a \Phi_1 \sum_{i=1}^3 \bar{q'_{idL}} q'_{idR} + b \Phi_1 \sum_{i \neq j}^3 \bar{q'_{idL}} q'_{jdR}  + c \Phi_2 \sum_{i,j,k} \epsilon_{ijk} \bar{q'_{idL}} q'_{jdR}/3 \nonumber \\ 
&+& a' \Phi_1 \sum_{i=1}^3 \bar{q'_{iuL}} q'_{iuR}  + b' \Phi_1 \sum_{i neq j} \bar{q'_{iuL}} q'_{juR} \nonumber \\
&+& c' \Phi_2 \sum_{i,j,k} \epsilon_{ijk} \bar{q'_{iuL}} q'_{juR}/3 + H.C. \nonumber ~~~~~~~~~~~~~~(4.4)
\end{eqnarray}

Therefore the coupling constants a, b, a' and b' correspond to the totally symmetric coupling and c, c' correspond to the totally antisymmetric.
From the above analysis, we find that there is another possible classification for the two Higgs doublets,
in which they transform according to the $\Gamma^3$.
 In the following, we are going to study this classification. \\

Now we assume that $\left( \begin{array}{cc} \Phi_1 \\ \Phi_2 \end{array}\right) $ transforms as $\Gamma^3$.
In order to construct S3 invariant Lagrangian, we should first extract the $\Gamma^3$ part in the product representation $\Gamma^6 \otimes \Gamma^6$. Then extract the singlet $\Gamma^1$ from $\Gamma^3 \otimes \Gamma^3$ \\

For the $\Gamma^3$ from $\Gamma^1 \otimes \Gamma^3$, $\bar{q'_L}$ is totally symmetric and takes the form $(\bar{q'_1}+\bar{q'_2}+ \bar{q'_3})_L$ while $q'_R$ is a doublet whose elements are linear combination of $q'_{1R},~q'_{2R}$ and $q'_{3R}$.
After some calculations, we find the doublet takes the form
 \begin{eqnarray}
\bar{q'_R}=(-2\bar{q'_1} +\bar{q'_2}+\bar{q'_3}, ~-\sqrt{3} \bar{q'_2}+\sqrt{3} \bar{q'_3})=(A,B) \nonumber ~~~~~~~~(4.5)
\end{eqnarray}

The $\Gamma^3$ from $\Gamma^3 \otimes \Gamma^1$ is similar.
The only difference between them is that the left (right)-handed quark fields form doublets (singlets) here. \\

The corresponding S3 invariant Lagrangian densities can be written as
 \begin{widetext}
\begin{eqnarray}
L^{1x3} &=&   (\bar{q'_1} + \bar{q'_2} +\bar{q'_3})_L q'_{dR} \left( \begin{array}{cc} 1 & 0 \\ 0 & 1 \end{array}\right) \left( \begin{array}{cc} \Phi_1 \\ \Phi_2 \end{array}\right) g_1  + (\bar{q'_1} + \bar{q'_2} +\bar{q'_3})_L q'_{uR} \left( \begin{array}{cc} 1 & 0 \\ 0 & 1 \end{array}\right) \left( \begin{array}{cc} \tilde{\Phi_1} \\ \tilde{\Phi_2} \end{array}\right) h_1 +H.C. \nonumber \\
&=& (\bar{q'_1},~ \bar{q'_2},~ \bar{q'_3})_L
          \left( \begin{array}{ccc}
         -2 g_1 \Phi_1 & g_1 \Phi_1 -\sqrt{3} g_1 \Phi_2 & g_1 \Phi_1 +\sqrt{3} g_1 \Phi_2 \\
           -2 g_1 \Phi_1 & g_1 \Phi_1 -\sqrt{3} g_1 \Phi_2 & g_1 \Phi_1 +\sqrt{3} g_1 \Phi_2 \\
            -2 g_1 \Phi_1 & g_1 \Phi_1 -\sqrt{3} g_1 \Phi_2 & g_1 \Phi_1 +\sqrt{3} g_1 \Phi_2
         \end{array}\right) \left( \begin{array}{ccc} d' \\ s' \\ b'  \end{array}\right)_R \nonumber \\
         &+& (\bar{q'_1},~ \bar{q'_2},~ \bar{q'_3})_L
           \left( \begin{array}{ccc}
          -2 h_1 \tilde{\Phi_1} & h_1 \tilde{\Phi_1}-\sqrt{3} h_1 \tilde{\Phi_2} & h_1 \tilde{\Phi_1} +\sqrt{3} h_1 \tilde{\Phi_2} \\
          -2 h_1 \tilde{\Phi_1} & h_1 \tilde{\Phi_1}-\sqrt{3} h_1 \tilde{\Phi_2} & h_1 \tilde{\Phi_1} +\sqrt{3} h_1 \tilde{\Phi_2} \\
        -2 h_1 \tilde{\Phi_1} & h_1 \tilde{\Phi_1}-\sqrt{3} h_1 \tilde{\Phi_2} & h_1 \tilde{\Phi_1} +\sqrt{3} h_1 \tilde{\Phi_2}
     \end{array}\right) \left( \begin{array}{ccc} u' \\ c' \\ t'  \end{array}\right)_R +H.C. \nonumber ~~~~~~~~~~~~~(4.6) \\
L^{3x1} &=& \bar{q'_L} (q'_1 +q'_2+q'_3)_{dR} \left( \begin{array}{cc} 1 & 0 \\ 0 & 1 \end{array}\right) \left( \begin{array}{cc} \Phi_1 \\ \Phi_2 \end{array}\right) g_2 + \bar{q'_L} (q'_1 +q'_2+q'_3)_{uR} \left( \begin{array}{cc} 1 & 0 \\ 0 & 1 \end{array}\right)
         \left( \begin{array}{cc} \tilde{\Phi_1} \\ \tilde{\Phi_2} \end{array}\right) h_2 +H.C. \nonumber \\
&=& (\bar{q'_1},~ \bar{q'_2},~ \bar{q'_3})_L
        \left( \begin{array}{ccc}
             -2 g_2 \Phi_1 & -2 g_2 \Phi_1 & -2 g_2 \Phi_1 \\
           g_2 \Phi_1 -\sqrt{3} g_2 \Phi_2 & g_2 \Phi_1 -\sqrt{3} g_2 \Phi_2 & g_2 \Phi_1 -\sqrt{3} g_2 \Phi_2 \\
                g_2 \Phi_1 +\sqrt{3} g_2 \Phi_2 & g_2 \Phi_1 +\sqrt{3} g_2 \Phi_2 & g_2 \Phi_1 +\sqrt{3} g_2 \Phi_2  \end{array}\right)
                \left( \begin{array}{ccc} d' \\ s' \\ b'  \end{array}\right)_R \nonumber \\
    &+& (\bar{q'_1},~ \bar{q'_2},~ \bar{q'_3})_L
        \left( \begin{array}{ccc}
     -2 h_2 \tilde{\Phi_1} & -2 h_2 \tilde{\Phi_1} & -2 h_2 \tilde{\Phi_1} \\
        h_2 \tilde{\Phi_1} -\sqrt{3} h_2 \tilde{\Phi_2} & h_2 \tilde{\Phi_1} -\sqrt{3} h_2 \tilde{\Phi_2} & h_2 \tilde{\Phi_1} -\sqrt{3} h_2 \tilde{\Phi_2} \\
                h_2 \tilde{\Phi_1} +\sqrt{3} h_2 \tilde{\Phi_2} & h_2 \tilde{\Phi_1} +\sqrt{3} h_2 \tilde{\Phi_2} & h_2 \tilde{\Phi_1} +\sqrt{3} h_2 \tilde{\Phi_2}  \end{array}\right)
                \left( \begin{array}{ccc} u' \\ c' \\ t'  \end{array}\right)_R  +H.C. \nonumber ~~~~~~~~(4.7)
\end{eqnarray}
 \end{widetext}

Now we consider the $\Gamma^3$ from $\Gamma^3 \otimes \Gamma^3$.
Let $\left( \begin{array}{cc} A \\ B \end{array}\right)$ and $\left( \begin{array}{cc} \bar{A} \\ \bar{B} \end{array}\right)$ be two S3 doulets.
It can be shown that $\left( \begin{array}{cc} \bar{A}A -\bar{B} B \\ -\Bar{A}B-\Bar{B}A \end{array}\right)$ is also an S3 doublet.
Therefore we have
 \begin{widetext}
\begin{eqnarray}
L^{3x3} &=&   (\bar{q'_1},~ \bar{q'_2} ,~\bar{q'_3})_L
 \left( \begin{array}{ccc}
 4g_3 \Phi_1 & -2g_3 \Phi_1-2\sqrt{3} g_3 \Phi_2 & -2g_3 \Phi_1+2\sqrt{3} g_3 \Phi_2 \\
 -2g_3 \Phi_1-2\sqrt{3} g_3 \Phi_2 & -2g_3 \Phi_1+2\sqrt{3} g_3 \Phi_2 & 4g_3 \Phi_1 \\
 -2g_3 \Phi_1+2\sqrt{3} g_3 \Phi_2 & 4g_3 \Phi_1 & -2g_3 \Phi_1-2\sqrt{3} g_3 \Phi_2 \end{array}\right)
 \left( \begin{array}{ccc} d'\\  s' \\ b' \end{array}\right)_R \nonumber \\
   &+& (\bar{q'_1} ,~ \bar{q'_2} ,~\bar{q'_3})_L  \left( \begin{array}{ccc}
   4h_3 \tilde{\Phi_1} & -2h_3\tilde{\Phi_1}-2 \sqrt{3} h_3 \tilde{\Phi_2} &  -2h_3\tilde{\Phi_1}+2 \sqrt{3} h_3 \tilde{\Phi_2} \\
   -2h_3\tilde{\Phi_1}-2 \sqrt{3} h_3 \tilde{\Phi_2} & -2h_3\tilde{\Phi_1}+2 \sqrt{3} h_3 \tilde{\Phi_2} & 4h_3 \tilde{\Phi_1} \\
    -2h_3\tilde{\Phi_1}+2 \sqrt{3} h_3 \tilde{\Phi_2} & 4h_3 \tilde{\Phi_1} & -2h_3\tilde{\Phi_1}-2 \sqrt{3} h_3 \tilde{\Phi_2} \end{array}\right)
    \left( \begin{array}{ccc} u' \\ c' \\ t' \end{array}\right) +H.C. \nonumber ~~~~~~~~(4.8)
\end{eqnarray}
 \end{widetext}

The total Lagrangian density is
 \begin{eqnarray}
L_{int}=L^{1x3} + L^{3x1} +L^{3x3} \nonumber ~~~~~~~~(4.9)
\end{eqnarray}

From the expressions for $L^{1x3}$, $L^{3x1} $ and $L^{3x3}$ we find that they are all traceless.
This result implies that the mass matrices are also traceless.
Therefore the sum of the quark masses is zero,
meaning that at least one of the quark masses is negative or all masses are zero, contrary to nature.
Therefore the classification in which the two Higgs-doublets form an S3 doublet and the quark fields belong to the three-dimensional representation $\Gamma^3$  is eliminated and should be modified. \\

\section{CONCLUSIONS AND DISCCUSIONS}

In this article, we have considered the two Higgs-doublet extension of the standard electroweak model and impose the permutation S3 symmetry on the Higgs sector.
We find that when we classify the quark fields to $\Gamma^6$, one Higgs doublet to $\Gamma^1$ and the other Higgs doublet to $\Gamma^2$,
we obtain a reasonable relation among the quark masses which predicts the mass of the top quark to be 39 Gev at 1 Gev renormalization scale.
We also find that according to this classification the flavour-changing neutral currents vanish automatically,
although each type of quark is coupled to both Higgs doublets.
Unfortunately, in the above classification no generation mixing and CP-violating phase show up.
If we classify the two Higgs doublets to $\Gamma^3$,
good results are not obtained. These effects suggest that in order to derive reasonable quark mass spectra and the weak mixing matrix,
we ought either to consider more Higgs doublets or to classify the right-handed quarks differently from the lefthanded one,
if we insist on the symmetry SU(2) x U(1) x S3 and the spontaneous breaking of S3.

\appendix
\section{THE VACUUM EXPECTATION VALUES OF THE TWO HIGGS DOUBLETS}

We denote the Higgs doublets as
\begin{eqnarray}
 \Phi_1 = \left( \begin{array}{cc} \Phi_1^+ \\ \Phi_1^0 \end{array}\right),~~
 \Phi_2 = \left( \begin{array}{cc} \Phi_2^+ \\ \Phi_2^0 \end{array}\right)
\end{eqnarray}

Then the most general potential invariant under SU(2) x U( 1) is
\begin{eqnarray}
 V(\Phi) &=& -\lambda_1 \Phi_1^{\dagger} \Phi_1-\lambda_2 \Phi_2^{\dagger} \Phi_2  +A(\Phi_1^{\dagger} \Phi_1)^2 \nonumber \\
 &+& B(\Phi_2^{\dagger} \Phi_2)^2  +C(\Phi_1^{\dagger} \Phi_1)(\Phi_2^{\dagger} \Phi_2) +D(\Phi_1^{\dagger} \Phi_2)(\Phi_2^{\dagger} \Phi_1) \nonumber \\
 &+& [(\Phi_1^{\dagger} \Phi_2)(E \Phi_1^{\dagger} \Phi_2 +F \Phi_1^{\dagger} \Phi_1 +G \Phi_2^{\dagger} \Phi_2)+H.C.]/2
\end{eqnarray}

We choose the vacuum expectation values as
\begin{eqnarray}
< \Phi_1 >= \left( \begin{array}{cc} \sigma \\ \rho_1 \end{array}\right)/\sqrt{2},~~
< \Phi_2 >= \left( \begin{array}{cc} 0 \\ \rho_2 \end{array}\right) e^{i \theta} /\sqrt{2}
\end{eqnarray}
in which $\rho_1$, $\rho_2$, $\sigma$, $\theta$ are real number and $\rho_1$, $\rho_2$, $\sigma >$ 0.
We substitute $< \Phi_1 >$ and $< \Phi_2 >$ into $V(\Phi)$
\begin{eqnarray}
 V(\Phi) &=& -\lambda_1 (\sigma^2 +\rho_1^2)/2 -\lambda_2 \rho_2^2 /2 + [A(\sigma^2 +\rho_1^2)^2 \nonumber \\
  &+& B \rho_2^4  + C \rho_2^2 (\sigma^2 +\rho_1^2) + (D-E) \rho_1^2 \rho_2^2]/4 \nonumber \\
  &+& E \rho_1^2 \rho_2^2 [(cos \theta -\Delta)^2-\Delta]/2 
 \end{eqnarray}
in which
\begin{eqnarray}
 \Delta = -{[F(\rho_1^2 +\sigma^2)+G \rho_2^2] \over {4E \rho_1 \rho_2}}
\end{eqnarray}

We take $cos \theta$, $\rho_1^2$, $\rho_2^2$ and $\sigma^2$ as four independent parameters.
The minimum conditions are
 \begin{widetext}
\begin{eqnarray}
 \partial V/\partial (\rho_1^2) &=& - {\lambda_1 \over 2} + A(\sigma^2 +\rho_1^2)/2 + C \rho_2^2 /4 + (D-E) \rho_2^2 /4 -{{F[G \rho_2^2 +F(\sigma^2 +\rho_1^2)]} \over {16E}} =0 \\
 \partial V/\partial (\rho_2^2) &=& - {\lambda_2 \over 2} + B \rho_2^2 /2 + C (\rho_1^2 +\sigma^2) /4 + (D-E) \rho_1^2 /4 -{{G[G \rho_2^2 +F(\sigma^2 +\rho_1^2)]} \over {16E}} =0 \\
 \partial V/\partial (\sigma^2) &=& - {\lambda_1 \over 2} + B \rho_2^2 /2 +  C (\rho_1^2 +\sigma^2)/4 -{F[G \rho_2^2 +F(\sigma^2 +\rho_1^2)]} / 16E =0 \\
 \partial V/\partial (cos\theta) &=& E \rho_1^2 \rho_2^2 ( cos \theta -\Delta)=0 \\
 \partial^2 V/\partial (\sigma^2)^2 &=& E \rho_1^2 \rho_2^2 > 0
\end{eqnarray}
 \end{widetext}

This result implies
\begin{eqnarray}
 cos\theta &=& A, \\
 E~>0
\end{eqnarray}
and
\begin{eqnarray}
{{\partial V} \over  {\partial (\sigma^2)}}={{\partial V} \over  {\partial (\rho_1^2)}}+{{(E-D)\rho_2^2} \over  4}
\end{eqnarray}

Case(1): (E-D) $>$ 0 \\
If ${{\partial V} / {\partial \rho_1^2}}=0$, ${{\partial V} / {\partial \sigma^2}}=(E-D)\rho_2^2/4 <0$ and V increases as $| \sigma^2 |$ increases.
So the minimum occurs at $\sigma^2=0$. \\

If ${{\partial V} / {\partial \sigma^2}}=0$, ${{\partial V} / {\partial \rho_1^2}}=(D-E)\rho_2^2/4 >0$ and V decreases as $\rho_1^2$ increases.
Thence V has no minimum in this case.
That is, ${{\partial V} / {\partial (\sigma^2)}}$ cannot vanish in this case. \\

Case(2): (E-D) $<$ 0 \\
If ${{\partial V} / {\partial \sigma^2}}=0$, ${{\partial V} / {\partial (\rho_1^2)}} >$ 0, and the minimum occurs at $\rho_1^2 =0$.
If ${{\partial V} / {\partial \rho_1^2}}=0$, ${{\partial V} / {\partial \sigma^2}} <$ 0, and no minimum occurs.
This result means that in this case ${{\partial V} / {\partial \rho_1^2}}$ cannot vanish and that we should take $\rho_1^2 =0$ for minimum V. \\

Case(3): (E-D)=0 \\
If (E - D) = 0, ${{\partial V} / {\partial \rho_1^2}}={{\partial V} / {\partial \sigma^2}}=0$.
 Minima exist for both $\rho_1^2$ and $\sigma^2$.
 This result implies that both $\rho_1$ and $\sigma$ cannot vanish for minimum V. \\

From the above discussion, we summarize as follows: \\
\begin{eqnarray}
(1)~If~(E-D)~>0, ~< \Phi_1 >=  \left( \begin{array}{cc} 0 \\ \rho_1 \end{array}\right)/\sqrt{2} \\
(2)~If~(E-D)~<0, ~< \Phi_1 >=  \left( \begin{array}{cc} \sigma \\ 0 \end{array}\right)/\sqrt{2} \\
(3)~If~(E-D)~=0, ~< \Phi_1 >=  \left( \begin{array}{cc} \sigma \\ \rho_1 \end{array}\right)/\sqrt{2}
\end{eqnarray}

In this report, we choose the first case, that is, E $>$ D and
\begin{eqnarray}
< \Phi_1 >=  \left( \begin{array}{cc} 0 \\ \rho_1 \end{array}\right)/\sqrt{2},
~~~< \Phi_2 >=  \left( \begin{array}{cc} 0 \\ \rho_2 \end{array}\right) e^{i\theta} /\sqrt{2}.
\end{eqnarray}

When we require the potential term V to be S3-invariant, there are two cases to denote the change of V. \\

Case (1): $\Phi_1$ is totally symmetric and $\Phi_2$ is totally antisymmetric. \\

In this case
\begin{widetext}
\begin{eqnarray}
 V(\Phi) &=& -\lambda_1 (\Phi_1^{\dagger} \Phi_1)-\lambda_2 (\Phi_2^{\dagger} \Phi_2)  +A(\Phi_1^{\dagger} \Phi_1)^2 + B(\Phi_2^{\dagger} \Phi_2)^2  +C(\Phi_1^{\dagger} \Phi_1)(\Phi_2^{\dagger} \Phi_2) +D(\Phi_1^{\dagger} \Phi_2)(\Phi_2^{\dagger} \Phi_1) \nonumber \\
  &+& [(\Phi_1^{\dagger} \Phi_2)(E \Phi_1^{\dagger} \Phi_2 +F \Phi_1^{\dagger} \Phi_1 +G \Phi_2^{\dagger} \Phi_2)+H.C.]/2 \nonumber \\
  &=&S_3 V S_3^{-1} \nonumber \\
 &=& -\lambda_1 (\Phi_1^{\dagger} \Phi_1)-\lambda_2 (\Phi_2^{\dagger} \Phi_2)  +A(\Phi_1^{\dagger} \Phi_1)^2 + B(\Phi_2^{\dagger} \Phi_2)^2  +C(\Phi_1^{\dagger} \Phi_1)(\Phi_2^{\dagger} \Phi_2) +D(\Phi_1^{\dagger} \Phi_2)(\Phi_2^{\dagger} \Phi_1) \nonumber \\
 &+& [(\Phi_1^{\dagger} \Phi_2)(E \Phi_1^{\dagger} \Phi_2 -F \Phi_1^{\dagger} \Phi_1 -G \Phi_2^{\dagger} \Phi_2)+H.C.]/2
\end{eqnarray}
\end{widetext}

We obtain
\begin{eqnarray}
F \Phi_1^{\dagger} \Phi_1 +G \Phi_2^{\dagger} \Phi_2 +H.C. = -F \Phi_1^{\dagger} \Phi_1 -G \Phi_2^{\dagger} \Phi_2 -H.C.
\end{eqnarray}
implying that F=G=0. \\

When we substitute this result into (A5), we find $cos \theta =0$.
This implies that $\theta$ must be $\pi /2$ or $-\pi /2$.
It agrees with the result in section 3. Furthermore, the results (A14) (A15) and (A16) are unchanged by this assumption. \\

Case(2): $\Phi_1$ and $\Phi_2$ construct a doublet $\left( \begin{array}{cc} \Phi_1 \\ \Phi_2 \end{array}\right)$ of S3. \\

In this case
\begin{widetext}
\begin{eqnarray}
V = {{-\lambda}\over 2} (\Phi_1^{\dagger},~\Phi_2^{\dagger})\left( \begin{array}{cc} \Phi_1 \\ \Phi_2 \end{array}\right)
  -{{\mu} \over 4} [(\Phi_1^{\dagger},~\Phi_2^{\dagger})\left( \begin{array}{cc} \Phi_1 \\ \Phi_2 \end{array}\right)]^2 = {{-\lambda}\over 2} (\Phi_1^{\dagger} \Phi_1 +\Phi_2^{\dagger} \Phi_2)
  -{{\mu} \over 4} [(\Phi_1^{\dagger} \Phi_1)^2+(\Phi_2^{\dagger} \Phi_2)^2+2(\Phi_1^{\dagger} \Phi_1)(\Phi_2^{\dagger} \Phi_2)]
\end{eqnarray}
\end{widetext}

When we compare this potential with the most general form of V in (A4) we find that $\lambda=\lambda_1=\lambda_2$, $\mu=A=B=C/2$ and $D=E=F=G=0$. \\

\end{document}